\documentclass[journal]{IEEEtran}
\usepackage{color}
\usepackage{cite}

\ifCLASSINFOpdf 
   \usepackage[pdftex]{graphicx}
   \graphicspath{{../pdf/}{../jpeg/}}
   \DeclareGraphicsExtensions{.pdf,.jpeg,.png}
\else
   \usepackage[dvips]{graphicx}
   \graphicspath{{../eps/}}
   \DeclareGraphicsExtensions{.eps}
\fi 
\usepackage{makecell,multirow,diagbox}
\usepackage{pifont}
\usepackage{enumerate}
\usepackage{amsfonts}
\usepackage{booktabs}
\usepackage{algorithmic}
\usepackage{algorithm}
\usepackage[cmex10]{amsmath}

\usepackage{mathabx}
\usepackage{mathrsfs}
\usepackage{subeqnarray}
\usepackage{cases}

\begin{document}

\title{Secure and Multi-Step Computation Offloading and Resource Allocation in Ultra-Dense Multi-Task NOMA-Enabled IoT Networks }

\author{Tianqing~Zhou,~
        Yanyan~Fu,~
        Dong~Qin,~
        Xuefang~Nie,~
        Nan~Jiang,~
        and Chunguo~Li~
\thanks{This work was supported by National Natural Science Foundation of China under Grant Nos. 62261020, 61861017, 62062034, 62001201, 62171119, 61861018, 61961020, 61862025 and 61963017,  National Key Research and Development Program of China under Grant No. 2020YFB1807201, Natural Science Foundation of Jiangxi Province of China under Grant Nos. 20212BAB202004, 20212BAB202004 and 20212BAB212001, Key Research and Development Plan of Jiangsu Province Grant No. BE2021013-3, Special 03 Project and 5G Project of Jiangxi Province under Grant No. 20203ABC03W07. The corresponding author is Chunguo Li.}
\thanks{T. Zhou, Y. Fu, X. Nie and N. Jiang are with the School of Information Engineering, East China Jiaotong University, Nanchang 330013, China (email: zhoutian930@163.com; fuyanyan3640@163.com; Xuefangnie@163.com; jiangnan1018@acm.org).}
\thanks{D. Qin is with School of Information Engineering, Nanchang University, Nanchang 330031, China (e-mail: qindong@seu.edu.cn).}
\thanks{C. Li is with School of Information Science and Engineering, Southeast University, Nanjing 210096, China (email: chunguoli@seu.edu.cn).}
}



\maketitle

\begin{abstract}
Ultra-dense networks are widely regarded as a promising solution to explosively growing applications of Internet-of-Things (IoT) mobile devices (IMDs). However, complicated and severe interferences need to be tackled properly in such networks. To this end, both orthogonal multiple access (OMA) and non-orthogonal multiple access (NOMA) are utilized at first. Then, in order to attain a goal of green and secure computation offloading, under the proportional allocation of computational resources and the constraints of latency and security cost, joint device association, channel selection, security service assignment, power control and computation offloading are done for minimizing the overall energy consumed by all IMDs. It is noteworthy that multi-step computation offloading is concentrated to balance the network loads and utilize computing resources fully. Since the finally formulated problem is in a nonlinear mixed-integer form, it may be very difficult to find its closed-form solution. To solve it, an improved whale optimization algorithm (IWOA) is designed. As for this algorithm, the convergence, computational complexity and parallel implementation are analyzed in detail. Simulation results show that the designed algorithm may achieve lower energy consumption than other existing algorithms under the constraints of latency and security cost.
\end{abstract}


\section{Introduction}\label{sec 1}
\IEEEPARstart{W}{i}th the staggering development of the mobile Internet of Things (IoT), a great many of new delay-sensitive and computing-intensive applications emerge, such as smart homes, virtual reality, augmented reality, autonomous driving, etc. \cite{JZhao2020Aug,LQian2021Jul,JZhao2019Aug}. Although the computing power of IoT mobile devices (IMDs) has achieved a qualitative leap, due to the limited computing resources and battery capacity, they cannot support these applications well \cite{FLi2022May,RZhang2022Apr,QZhu2019Sep,JZhao2019Dec}. To address such an issue, mobile edge computing (MEC) is widely regarded as a promising option, which provides a large number of computing resources for IMDs (users) at the edge of networks. In MEC networks, any task of IMDs (users) can be partially or completely offloaded to some neighboring edge servers for computing. Evidently, through such an operation, the workloads and energy consumption of users may be reduced greatly.
\par
In order to further shorten the distance between users and computing centers, ultra-dense networks are widely advocated and have attracted increasing attention, where base stations (BSs) are equipped with MEC servers \cite{TZhou2021Dec}. Through the deployment of ultra-dense BSs, the service coverage can be enhanced greatly, and the uplink transmission power of users may be reduced significantly. However, such a deployment often results in complicated and severe network interferences. In addition, during the computation offloading, offloaded tasks are vulnerable to malicious attacks. To attain the goal of secure communications, some additional computation overheads yield for secure preventive services, resulting in extra computation latency and energy consumption.
\par
It is evident that the design of secure and green offloading mechanisms is an important topic in ultra-dense networks. Specifically, under the limited network resources, the central issue remains how to protect users' data, mitigate network interferences and reduce users' energy consumption in such networks.
\subsection{Related Work}
In wireless networks, although spectrum sharing is beneficial to improving spectrum utilization, it will inevitably incur severe interferences within and between regions. It means that a reasonable resource management strategy needs to be introduced, especially for ultra-dense networks. To this end, some relevant efforts have been made as follows. In \cite{FGuo2018Dec}, joint spectrum, power, computation offloading decisions and resource allocation were optimized to minimize the energy consumed by users in densely deployed small cell networks. Such work considered distinct channels for macro BSs (MBS) and small BSs (SBSs), but let users utilize the same channels at some BS. In \cite{YLi2022Sep}, offloading decisions, transmission duration and computing rate were jointly optimized to minimize the overall delay of tasks under both non-orthogonal multiple access (NOMA) and orthogonal frequency division multiple access (OFDMA). In \cite{LLi2021Mar}, Li \textit{et al.} jointly optimized uplink transmission power, offloading decisions and weight coefficients of delay and energy consumption to minimize energy consumption for ultra-dense networks with NOMA and time division multiple access (TDMA). In \cite{YLu2022Sep}, Lu \textit{et al.} jointly optimized task offloading, BS selection, channel and computing resource allocation to minimize the total system cost consisting of delay and energy consumption caused by users and BSs for ultra-dense networks with OFDMA.
\par
At the same time, multi-task offloading has attracted more and more attention in recent years. Some related work has been done as follows. In \cite{MHChen2018Oct}, joint task offloading decisions and bandwidth allocation were considered to minimize total system cost defined as the weighted sum of energy consumption and task delay. In such work, each user has multiple independent tasks. In \cite{MSun2020Oct}, joint multi-task offloading decisions, computing and spectral resource allocation were optimized to minimize task latency while guaranteeing the energy available to the users. Such an investigation was made under the user-assisted MEC system. In \cite{YWu2020Jul}, offloaded workloads and local computing rates were jointly optimized to minimize the weighted sum of energy consumed by NOMA transmission and local execution of smart terminals for a multi-task NOMA system. In \cite{HZhang2021Feb}, the amount of offloaded data was optimized to minimize task delay for a multi-server and multi-task scenario. In \cite{JChen2022Jun}, the computation offloading was performed to minimize the average energy-time cost of all users for a MEC system with multiple dependent tasks. In \cite{JBi2022Oct}, joint resource allocation and partial computation offloading were performed to minimize system energy consumption for heterogeneous edge networks with multiple separable tasks. In \cite{HTang2022Jun}, joint multi-task offloading and resource allocation were executed to minimize the weighted sum of delay and energy consumption under task-overflowed situations.
\par
It is easy to find that aforementioned one-step computation offloading cannot utilize computing resources well, and computation delay may increase with the number of tasks significantly. To fully utilize these resources in networks, especially in ultra-dense networks, multi-step computation offloading has been regarded as a good option. So far, multi-step computation offloading was rarely studied and is still an open topic. Some existing efforts made towards it can be listed as follows. In \cite{YDai2018Dec}, joint user association, multi-step offloading decision, power and computation resources were optimized to minimize network-wide energy consumption for ultra-dense multi-task networks under users' latency constraints. In \cite{TZhou2022Oct}, joint device association, multi-step computation offloading and resource allocation were performed to minimize the network-wide energy consumption for ultra-dense multi-task networks under OFMDA and proportional computing resource allocation. In \cite{HZhang2022Aug}, joint device association, channel allocation and multi-part collaborative offloading were optimized to minimize the average delay under the affordable cost of network operators.
\par
Although computation offloading can reduce energy consumed by mobile terminals and task delay greatly, offloaded data is vulnerable to malicious attacks. In view of this, secure computation offloading has attracted increasing attention.
\par
Although computation offloading can reduce energy consumed by mobile terminals and task delay greatly, offloaded data is vulnerable to malicious attacks. In view of this, secure computation offloading has attracted increasing attention. There exists some related work listed as follows. In \cite{SHan2019Jun}, Han \textit{et al.} jointly optimized computing and communicational resources to maximize the secrecy energy efficiency of computation offloading in a NOMA system. In \cite{XHe2020jun}, He \textit{et al.} jointly optimized offloading ratio and uplink transmission power to energy-plus-payment cost, where some jamming signals broadcasted by edge servers were used for impeding eavesdropping. In \cite{JBWang2020Aug}, Wang \textit{et al.} jointly optimized uplink transmission power, offloading timeslots, task allocation and local processing frequency to minimize system energy consumption under physical layer security techniques. In \cite{WWu2020Jan}, Wu \textit{et al.} jointly optimized task partition, power allocation, codeword transmission rate and confidential data rate to minimize the weighted sum of energy consumption under physical layer security and NOMA techniques. In \cite{YBai2020Jun}, Bai \textit{et al.} jointly optimized offloading and attacking decisions to maximize the expected reward of the edge system, which involves both the service delay and security risks. In \cite{{SLiu2022Jun}}, Liu \textit{et al.} jointly optimized task partition, uplink transmission power and offloading rate to maximize the requirement satisfaction of all users, which is quantized as a combination of delay, energy consumption and security decisions.
\par
It is evident that the above-mentioned work concentrated on physical-layer assisted secure offloading mechanisms. When many attackers cooperate with each other, such mechanisms cannot guarantee the task security well. In view of this, some secure offloading schemes based on cryptographic algorithms have attracted increasing attention. In \cite{IElgendy2019}, Elgendy \textit{et al.} jointly optimized security decisions, resource allocation and computation offloading to minimize the energy consumption and delay of the entire system. In addition, they also jointly optimized security decisions, offloading policy, task compression and resource allocation to minimize the weighted sum of energy consumption for a multi-task MEC system \cite{IAElgendy2020Dec}. After that, according to the execution time, energy consumption, CPU and memory usage, the computation offloading was dynamically performed in \cite{IAElgendy2021Jan}, where a new security layer was added to protect the transferred data in the cloud. In \cite{MZahed2020}, Zahed \textit{et al.} jointly optimized security service assignment, cooperative task offloading and caching to minimize the total system cost quantized as a combination of security breach cost and energy consumption.
\par
Among the above-mentioned efforts, few concentrate on the design of secure multi-task multi-step computing offloading mechanisms, especially for ultra-dense networks. In addition, most of them utilize the frequency spectrum of ultra-dense networks in a pure OFDMA or full-frequency reusing manner. However, since such manners may result in low spectrum efficiency or severe network interferences, they may be unreasonable and impractical for such networks.
\subsection{Contributions and Organization}
Unlike most efforts, in ultra-dense multi-task IoT networks, we first consider the BS clustering, OFDMA and NOMA to mitigate network interferences and improve frequency spectrum utilization. After that, we try to develop a secure and green computation offloading scheme to minimize the energy consumed by IMDs under constraints of latency and secure costs, which jointly optimizes the device association, channel selection, task partition, security service assignment, uplink transmission power and computing resource allocation. Specifically, the main work and contributions of this paper can be summarized as follows:
\begin{enumerate}
\item \textit{Joint BS Clustering, OFDMA and NOMA Used for Ultra-Dense IoT Networks:} To mitigate the complicated and severe interferences, and improve frequency spectrum utilization, we consider the following operations for ultra-dense IoT networks. At first, SBSs are first divided into several clusters using K-means according to their physical positions. Secondly, the whole system frequency band is cut into two parts used by MBS and SBS separately. Thirdly, we let each cluster own some orthogonal subchannels (frequency bands), but different clusters have distinct subchannels. At last, we consider that IMDs served by SBSs in the same cluster perform uplink transmission in a NOMA manner, but ones served by MBSs utilize frequency bands equally. As far as we know, such joint BS Clustering, OFDMA and NOMA should be a new investigation for ultra-dense IoT networks.
\item \textit{Secure Multi-Step Multi-Task Computation Offloading in Ultra-Dense IoT Networks:} In ultra-dense multi-task IoT networks, we consider secure one-step and two-step computation offloading. In a one-step manner, a part of any task of an IMD is offloaded to the associated MBS. In a two-step manner, a part of any task of an IMD is first offloaded to the associated SBS, and then a part of the partial task received by the SBS is further offloaded to a nearby MBS. To attain the goal of secure green communications, any offloaded part needs to be encrypted, and receivers decrypt received parts. To the best of our knowledge, such secure multi-step multi-task computation offloading should be a new topic in ultra-dense IoT networks.
\item \textit{ Problem Formulation of Secure Multi-Step Multi-Task Computation Offloading in Ultra-Dense IoT Networks:} To achieve the goal of green and secure computation offloading in ultra-dense multi-task IoT Networks, under joint BS clustering, OFDMA and NOMA, proportional allocation of computational resources, and the constraints of latency and security costs, we jointly optimize device association, channel selection, security service assignment, power control and computation computing resources to minimize the total energy consumed by all IMDs. Evidently, it should be a new formulation.
\item \textit{ Design Algorithm to Solve the Formulated Problem:} Considering that the formulated problem is in a nonlinear mixed-integer form, we design an improved whale optimization algorithm (IWOA). Specifically, we first improve conventional WOA by changing parameters, settings and rules in three phases consisting of searching for prey, shrinking encirclement and bubble-net attacking. In addition, we replace the current best agent with the historically best agent, and search for prey in the nearby area of the historically best agent.
\item \textit{ Analyses of Convergence, Computation Complexity and Simulation:} As for the designed algorithm in this paper, we provide some detailed analyses of the convergence and computation complexity. At last, we investigate its effectiveness by introducing other existing algorithms for comparison in the simulation.
\end{enumerate}
\par
The rest of this article is organized as follows. The second section introduces the system model, including the network model, communication model, computing model, security model, and multi-task model; the third section gives the optimization problem formulation of minimizing the energy consumption of the whole network under the constraints of IMDs delay and the total cost of security vulnerabilities;
Section IV develops the IWOA-IPSO algorithm to solve the stated problem; Section V provides a detailed algorithm analysis, including three parts: convergence, computational complexity, and parallel implementation; Section VI presents the simulation results and analysis; Section VII presents conclusions and directions for further research in the future.
\begin{figure}[!t]
	\centering
	\centerline{\includegraphics[width=3.6in]{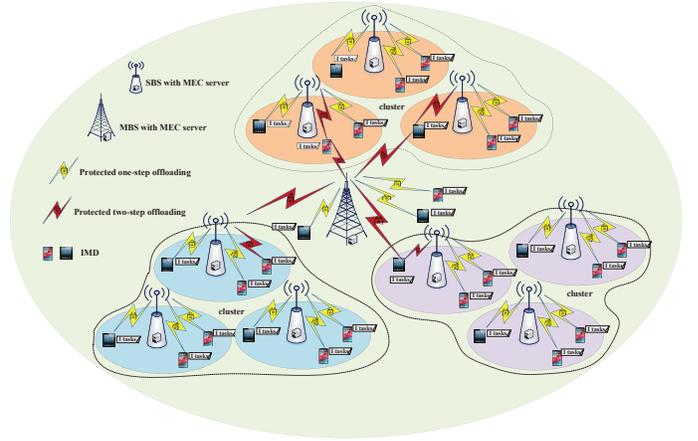}}
	\caption{Ultra-dense multi-task IoT networks with secure multi-step offloading.}
	\label{fig1}
\end{figure}
\section{SYSTEM MODEL}\label{sec 2}
In this section, network, communication, security and computation models are given in detail.
\subsection{Network Model}
In this paper, we concentrate on ultra-dense multi-task IoT networks with secure multi-step offloading, which is illustrated in Fig.\ref{fig1}. In such networks, the number of SBSs is greater than or equal to the one of IMDs; each BS is equipped with a MEC server; all SBSs are connected to nearby MBS via wired links; each IMD has $K$ independent delay-sensitive and computing-intensive tasks to execute within a security breach cost and a specific deadline. Without loss of generality, we consider that there exists an MBS and $\bar{S}$ SBSs in Fig.\ref{fig1}, where $\bar{S}$ SBSs are indexed from 1 to $\bar{S}$ in the set $\bar{\mathcal{S}}=\big\{ 1,2,\cdots ,\bar{S} \big\}$; the index of MBS is 0; $\mathcal{S}=\bar{\mathcal{S}}\cup \left\{ 0 \right\}$ represents the set of all BSs; $U$ IMDs are indexed from 1 to $U$ in the set $\mathcal{U}=\{1,2,\cdots ,U\}$; the tasks of each IMDs are indexed from 1 to $K$ in the set $\mathcal{K}=\{1,2,\cdots ,K\}$.
\par
In Fig.\ref{fig1}, when an IMD is associated with some SBS, a part of any task of this IMD is offloaded to such BS after encrypting. This BS first decrypts it and then transmits its part to nearby MBS after encrypting. Significantly, the associated SBS executes the remaining part, and MBS calculates the received part after decrypting. When an IMD is associated with some MBS, a part of any task of this IMD is offloaded to such BS after encrypting. This BS executes it after decrypting. Evidently, SBSs concentrate on secure two-step offloading, but MBSs adopt secure one-step offloading.
\par
To mitigate cross-tier interferences, the system frequency band is cut into two parts used by MBS and SBS separately, where the widths of them are $\eta {\varpi}$ and $(1-\eta ) {\varpi}$ respectively; $\varpi$ is the width of the system frequency band; $0\leq\eta\leq1$ is the band division factor. To further improve the spectrum efficiency, SBSs are divided into $W$ clusters using K-means according to their physical positions, where each cluster has $N$ orthogonal subchannels (frequency bands) used by SBSs in this cluster, and IMDs associated with these SBSs can utilize the same subchannel through a NOMA manner; $N$ subchannels are indexed from 1 to $N$ in the set $\mathcal{N}$; $N=\text{round}({(1-\eta ){\varpi}}/{\left( {\omega}{M} \right)})$, round( ) is a rounding function, and ${\omega}$ is the bandwidth of a subchannel. Significantly, IMDs associated with some MBS utilize frequency band $\eta {\varpi}$ equally.
\subsection{Communication Model}
Under the aforementioned resource utilization manner, there just exist intra-cluster interferences exist for any task of IMDs. In view of this, the uplink data rate of IMD $i$ associated with SBS $s\in \bar{\mathcal{S}}$ on subchannel $n$ can be given by
\begin{equation}\label{eq1}
\left\{ \begin{aligned}
  & {{R}_{i,s,n}}={\omega}{{\log }_{2}}\Big( 1+\frac{{{p}_{i}}{{\hbar}_{i,s}}}{\sum\nolimits_{u\in {{\mathcal{Q}}_{i,s,n}}}{{{p}_{u}}{{\hbar}_{u,s}}}+{{\sigma }^{2}}} \Big), \\
 & {{\mathcal{Q}}_{i,s,n}}=\left\{ i\in \mathcal{U}\right\}\backslash \left\{ i=u \right\}: \\
 &\ \ \ \ \ \  \ \ {{\hbar}_{u,s}}\le {{\hbar}_{i,s}};{{a}_{u}}={{a}_{i}}=n;{{b}_{u}},{{b}_{i}}\in {{\mathcal{W}}_{s}}, \\
\end{aligned} \right.
\end{equation}
where ${{a}_{i}}$ and ${{b}_{i}}$ are the channel and SBS indices selected by IMD $i$ respectively; ${{\hbar}_{i,s}}$ is channel gain between IMD $i$ and BS $s$; $p_i$ is the transmission power of IMD $i$; $\sigma^2$ is the noise power; ${{\mathcal{W}}_{s}}$ denotes the cluster that SBS $s$ belongs to.
\par
Since IMDs associated with an MBS utilize frequency bands equally, and these bands are different from the ones used by SBSs, there are no intra-tier and cross-tier interferences. Considering that IMDs often transmit tasks one by one on some channel, we can assume that each MBS has $N$ virtual subchannels, which correspond to only one channel in reality. That is to say, any IMD can use one of them to transmit a task at some time slot, which means that such an IMD utilizes a real channel to do it. Based on this, the uplink data rate of IMD $i$ associated with MBS 0 on subchannel $n$ can be given by
\begin{equation}\label{eq2}
    {{R}_{i,0,n}}=\eta {{\varpi}}{{\big( \sum\nolimits_{u\in \mathcal{U}}{{{x}_{u,0}}} \big)}^{-1}}{{\log}_{2}}\left( 1+{{{p}_{i}}{{\hbar}_{i,0}}}/{{{\sigma }^{2}}} \right),
\end{equation}
where $\sum\nolimits_{u\in \mathcal{U}}{{{x}_{u,0}}}$ is the number of IMDs associated with MBS 0; ${{x}_{i,s}}$ denotes the association index of IMD $i$ at BS $s$; ${{x}_{i,s}}=1$ if IMD $i$ is associated with the BS $s$; otherwise, ${{x}_{i,s}}=0$.
\subsection{Security Model}
In the reality, offloaded tasks often have different security requirements. However, they may be vulnerable to malicious attacks, eavesdropping, and spoofing. To tackle such an issue, data encryption and decryption are widely regarded as promising solutions, which utilize different cryptographic algorithms. As revealed in \cite{MZahed2020}, as the strength and robustness of security protection algorithms increase, the energy and latency overhead increase significantly. In addition, these preventive measures prevent security breaches completely. Therefore, quantifying security risks is an important topic in the design of secure offloading strategies.
\par
To guarantee secure offloading, offloaded tasks are encrypted and decrypted using different cryptographic algorithms in this paper. When an IMD is associated with some SBS, a part of any task of this IMD is offloaded to such BS after encrypting. This BS first decrypts it and then transmits its part to nearby MBS after encrypting. Significantly, the associated SBS executes the remaining part, and MBS calculates the received part after decrypting. When an IMD is associated with some MBS, a part of any task of this IMD is offloaded to such BS after encrypting. This BS executes it after decrypting.
\par
As we know, distinct cryptographic algorithms correspond to distinguishable security levels. We assume that security protection levels are indexed from ${{v}_{1}}$ to ${{v}_{L}}$ in the set $\mathcal{V}=\{{{v}_{1}},{{v}_{2}},\cdots ,{{v}_{l}},\cdots ,{{v}_{L}}\}$, and ${{v}_{l}}=l$ represents protection level (robustness) of the cryptographic algorithm $l$. In addition, the computation capacities of cryptographic algorithm $l$ are $\bar{\theta}_l$ (in CPU cycles/bit) and $\hat{\theta}_l$ (in CPU cycles/bit) for encrypting and decrypting one bit, and the corresponding energy consumptions are assumed to be the same, i.e., $\tilde{\theta}_l$ (in mJ /bit). Significantly, $\boldsymbol{\bar{\theta}}=\{\bar{\theta}_l,\forall l \in \mathcal{L}\}$, $\boldsymbol{\hat{\theta}}=\{\hat{\theta}_l,\forall l \in \mathcal{L}\}$ and $\boldsymbol{\tilde{\theta}}=\{\tilde{\theta}_l,\forall l \in \mathcal{L}\}$.
\par
When task $k$ of IMD $i$ adopts cryptographic algorithm $l$ to offload its parts securely, its failure probability \cite{WJiang2015Aug} can be given by
\begin{equation}\label{eq3}
\bar{p}_{i,k,l}=\left\{\begin{array}{l}
		1-e^{-\nu_{i,k}\left(\rho_{i,k}-v_l\right)}, \text { if } v_l<\rho_{i,k}, \\
		0, \text { otherwise},
	\end{array}\right.
\end{equation}
where $\nu_{i,k}$ is the security risk coefficient of task $k$ of IMD $i$; $\rho_{i,k}$ is the expected security level of task $k$ of IMD $i$. As revealed in \eqref{eq3}, cryptographic algorithm $l$ successfully protects task $m$ of IMD $i$ if its security level is greater or equal to the expected one. Otherwise, algorithm $l$ fails in protecting such a task.
\par
The security breach cost \cite{WJiang2015Aug} of task $k$ of IMD $i$ can be given by
\begin{equation}\label{eq4}
\varphi_{i,k}=\sum\nolimits_{s\in \mathcal{S}}{\sum\nolimits_{l \in \mathcal{L}} {\lambda_{k}}{{x}_{i,s}}{{y}_{i,k,l}} \bar{p}_{i,k,l}},
\end{equation}
where $\lambda_{k}$ is the finance loss (in \$) of task $k$ if it fails; ${{y}_{i,k,l}}$ is the security decision index of the task $k$ of IMD $i$, ${{y}_{i,k,l}}=1$ if cryptographic algorithm $l$ is selected for tackling task $k$ of IMD $i$, 0 otherwise. Then, overall security breach cost of IMD $i$ can be given by
\begin{equation}\label{eq5}
\psi_i=\sum\nolimits_{k\in \mathcal{K}} \varphi_{i,k}=\sum\nolimits_{s\in \mathcal{S}}\sum\nolimits_{k\in \mathcal{K}} \sum\nolimits_{l \in \mathcal{L}} {\lambda_{k}}{{x}_{i,s}}{{y}_{i,k,l}}\bar{p}_{i,k,l},
\end{equation}
where ${{\mu }_{i,k}}$ is the cost incurred by failure security protection of task $k$ of IMD $i$.
\subsection{Computational Model}
Task $k$ of IMD $i$ is denoted as ${{\mathcal{D}}_{i,k}}\triangleq \left( {{d}_{i,k}},{{c}_{i,k}},{\tau}_{i}^{\max},\rho_{i,k} \right)$, where ${{d}_{i,k}}$ represents the data size of task $k$ of IMD $i$; ${{c}_{i,k}}$ is the number of CPU cycles used to calculate one bit of task $k$ of IMD $i$; ${\tau}_{i}^{\max}$ is the deadline of IMD $i$.
\par
\textit{\textbf{1) Local computation: }}When IMD $i$ is associated with BS $s$, the size of locally processed data of task $k$ of IMD $i$ is ${{d}_{i,k}}-{{\bar{d}}_{i,s,k}}$, where ${{\bar{d}}_{i,s,k}}$ is the data size of task $k$ offloaded from IMD $i$ to BS $s$. In addition, the local executing time ${\tau}_{i,s,k}^{LOC}$ used for processing the task $k$ of IMD $i$ associated with BS $s$ can be given by
\begin{equation}\label{eq6}
	{\tau}_{_{i,s,k}}^{LOC}=\frac{\left( {{d}_{i,k}}-{{{\bar{d}}}_{i,s,k}} \right){{c}_{i,k}}}{{{f}_{i}^{UE}}}+\sum\limits_{l\in \mathcal{L}}{\frac{{{y}_{i,k,l}}{{{\bar{\theta }}}_{l}}{{{\bar{d}}}_{i,s,k}}}{{{f}_{i}^{UE}}}},
\end{equation}
where ${{f}_{i}^{UE}}$ is the computing capability (capacity) of IMD $i$; the two items on the right side of \eqref{eq6} are the computing time and encrypting time respectively.
\par
\textit{\textbf{2) Offloading to SBS: }} When IMD $i$ is associated with SBS $s$, the following steps need to be executed for any task $k$. At first, the part ${{{\bar{d}}}_{i,s,k}}$ of ${{d}_{i,k}}$ is offloaded from IMD $i$ to SBS $s$ after encrypting. Secondly, SBS $s$ decrypts ${{{\bar{d}}}_{i,s,k}}$, and then executes ${{{\bar{d}}}_{i,s,k}}-{{{\hat{d}}}_{i,s,k}}$. Thirdly, the part ${{{\hat{d}}}_{i,s,k}}$ of ${{{\bar{d}}}_{i,s,k}}$ is offloaded from SBS $s$ to nearby MBS after encrypting. Fourthly, MBS executes ${{{\hat{d}}}_{i,s,k}}$ after decrypting. Consequently, the remote time ${\tau}_{i,s,k}^{BS}$ used for processing the task $k$ of IMD $i$ associated with SBS $s$ can be given by
\begin{equation}\label{eq7}
	\begin{aligned}
		{\tau}_{i,s,k}^{BS}=&\sum\limits_{n\in \mathcal{N}}\frac{{z}_{i,n}{{{\bar{d}}}_{i,s,k}}}{{{R}_{i,s,n}}}+\frac{\big( {{{\bar{d}}}_{i,s,k}}-{{{\hat{d}}}_{i,s,k}} \big){{c}_{i,k}}}{{{{\bar{f}}}_{i,s,k}}}+\frac{{{{\hat{d}}}_{i,s,k}}}{{{R}_{0}}}\\
&+\frac{{{{\hat{d}}}_{i,s,k}}{{c}_{i,k}}}{{{{\bar{f}}}_{i,0,k}}}+\sum\nolimits_{l\in \mathcal{L}}{\frac{{{y}_{i,k,l}}{{{\hat{\theta }}}_{l}}{{{\bar{d}}}_{i,s,k}}}{{{{\bar{f}}}_{i,s,k}}}}\\
&+\sum\nolimits_{l\in \mathcal{L}}{\frac{{{y}_{i,k,l}}{{{\bar{\theta }}}_{l}}{{{\hat{d}}}_{i,s,k}}}{{{{\bar{f}}}_{i,s,k}}}}+\sum\nolimits_{l\in \mathcal{L}}{\frac{{{y}_{i,k,l}}{{{\hat{\theta }}}_{l}}{{{\hat{d}}}_{i,s,k}}}{{{{\bar{f}}}_{i,0,k}}}}, \\
	\end{aligned}
\end{equation}
where ${{z}_{i,n}}$ denotes the association decision of IMD $i$ on subchannel $n$; ${{z}_{i,n}}=1$ if IMD $i$ selects subchannel $n$, ${{z}_{i,n}}=0$ otherwise. ${{R}_{0}}$ is the wired backhaul rate between SBS and MBS; ${{\bar{f}}_{i,s,k}}$ is the computing capability allocated to task $k$ of IMD $i$ by SBS $s$; on the right side of \eqref{eq7}, the first four items are the time used for uploading ${{{\bar{d}}}_{i,s,k}}$ from IMD $i$ to SBS $s$, the one used for computing ${{{\bar{d}}}_{i,s,k}}-{{{\hat{d}}}_{i,s,k}}$ at SBS $s$, the one used for uploading ${{{\hat{d}}}_{i,s,k}}$ from SBS $s$ to nearby MBS, and the one used for computing ${{{\hat{d}}}_{i,s,k}}$ at MBS, respectively. The last three items are the time used for decrypting ${{{\bar{d}}}_{i,s,k}}$ at SBS $s$, the one used for encrypting ${{{\hat{d}}}_{i,s,k}}$, and the one used for decrypting ${{{\hat{d}}}_{i,s,k}}$, respectively.
\par
According to the ratio of CPU cycles used for tackling task $k$ of IMD $i$ to total utilized cycles at associated BS $s$, the computing capability of BS $s$ is allocated to the computing and secure operations of such a task. Specifically, when IMD $i$ is associated with SBS $s$, the computing capability $\bar{f}_{i,s,k}$ assigned to $k$ of IMD $i$ by SBS $s$ can be given by
\begin{equation}\label{eq8}
{{{\bar{f}}}_{i,s,k}}=\frac{f_{s}^{BS}\left({\Gamma}_{i,s,k}+\sum\nolimits_{l\in \mathcal{L}}{{{y}_{i,k,l}}\bar{\Gamma}_{i,s,k,l}} \right)}{\sum\nolimits_{u\in \mathcal{U}}{\sum\nolimits_{j\in \mathcal{K}}{{{x}_{u,s}}\left({\Gamma}_{u,s,j}+ \sum\nolimits_{l\in \mathcal{L}}{{{y}_{u,j,l}}\bar{\Gamma}_{u,s,j,l}} \right)}}},
\end{equation}
\begin{equation}\label{eq9}
\left\{ \begin{aligned}
  & {\Gamma}_{i,s,k}=\big( {{{\bar{d}}}_{i,s,k}}-{{{\hat{d}}}_{i,s,k}} \big){{c}_{i,k}}, \\
 & \bar{\Gamma}_{i,s,k,l}={{{\hat{\theta }}}_{l}}{{{\bar{d}}}_{i,s,k}}+{{{\bar{\theta }}}_{l}}{{{\hat{d}}}_{i,s,k}}, \\
\end{aligned} \right.
\end{equation}
where $f_{s}^{BS}$ represents total computing capability of SBS $s$; ${\Gamma}_{i,s,k}$ is the CPU cycles used for processing ${{{\bar{d}}}_{i,s,k}}-{{{\hat{d}}}_{i,s,k}}$; $\bar{\Gamma}_{i,s,k,l}$ is the CPU cycles used for decrypting ${{{\bar{d}}}_{i,s,k}}$ and encrypting ${{{\hat{d}}}_{i,s,k}}$.
\par
Since IMDs associated with SBSs can further offload partial tasks to nearby MBSs for processing, and ones associated with MBSs can directly upload tasks to these BSs for execution, the data processed at any MBS should include the following two parts. Consequently, the CPU cycles used for computing and decrypting the data offloaded from SBSs to this MBS selected by IMDs, which is given by $\sum\nolimits_{u\in\mathcal{U}}{\sum\nolimits_{s\in \bar{\mathcal{S}}}{{{x}_{u,s}}\sum\nolimits_{j\in \mathcal{K}}{{\Upsilon}_{u,s,j}}}}$, where ${\Upsilon}_{u,s,j} ={{{\hat{d}}}_{u,s,j}}{{c}_{u,j}}+\sum\nolimits_{l\in\mathcal{L}}{{{y}_{u,j,l}}{{{\hat{\theta }}}_{l}}{{{\hat{d}}}_{u,s,j}}}$. In addition, the CPU cycles used for computing and decrypting the data offloaded from IMDs to MBS selected by them, which is given by $\sum\nolimits_{u\in\mathcal{U}}{{{x}_{u,0}}\sum\nolimits_{j\in \mathcal{K}}{\bar{\Upsilon}_{u,0,j} }}$, where $\bar{\Upsilon}_{u,0,j} ={{{\bar{d}}}_{u,0,j}}{{c}_{u,j}}+\sum\nolimits_{l\in \mathcal{L}}{{{y}_{u,j,l}}{{{\hat{\theta }}}_{l}}{{{\bar{d}}}_{u,0,j}}}$. Under the proportional computing allocation mentioned previously, when IMD $i$ is associated with MBS 0, the computing capability $\bar{f}_{i,0,k}$ assigned to $k$ of IMD $i$ by MBS 0 can be given by
\begin{equation}\label{eq10}
	{{\bar{f}}_{i,0,k}}=\frac{f_{0}^{BS}(\sum\nolimits_{s\in \bar{\mathcal{S}}}{{{x}_{i,s}}{{\Upsilon }_{u,s,j}}}+{{x}_{i,0}}{{{\bar{\Upsilon }}}_{u,0,j}})}{\sum\nolimits_{u\in \mathcal{U}}{\sum\nolimits_{j\in \mathcal{K}}{\left( \sum\nolimits_{s\in \bar{\mathcal{S}}}{{{x}_{u,s}}{{\Upsilon }_{u,s,j}}}+{{x}_{u,0}}{{{\bar{\Upsilon }}}_{u,0,j}} \right)}}}.
\end{equation}
\par
\textit{\textbf{3) Offloading to MBS: }} When IMD $i$ is associated with MBS 0, the following steps need to be executed for any task $k$. At first, the part ${{{\bar{d}}}_{i,s,k}}$ of ${{d}_{i,k}}$ is offloaded from IMD $i$ to MBS 0 after encrypting. Secondly, MBS 0 decrypts ${{{\bar{d}}}_{i,s,k}}$, and then executes it. Consequently, the remote time ${\tau}_{i,0,k}^{BS}$ used for processing the task $k$ of IMD $i$ associated with MBS 0 can be given by
\begin{equation}\label{eq11}
    {\tau}_{i,0,k}^{BS}=\sum\limits_{n\in \mathcal{N}}\frac{{z}_{i,n}{{{\bar{d}}}_{i,0,k}}}{{{R}_{i,0,n}}}+\frac{{{{\bar{d}}}_{i,0,k}}{{c}_{i,k}}}{{{{\bar{f}}}_{i,0,k}}}+\sum\limits_{l\in \mathcal{L}}{\frac{{{y}_{i,k,l}}{{{\bar{d}}}_{i,0,k}}{{{\hat{\theta }}}_{l}}}{{{{\bar{f}}}_{i,0,k}}}},
\end{equation}
where the items on the right side of \eqref{eq11} represent the time used for uploading ${{{\bar{d}}}_{i,s,k}}$ from IMD $i$ to MBS 0, the one used for computing ${{{\bar{d}}}_{i,s,k}}$ at MBS 0, and the one used for decrypting ${{{\bar{d}}}_{i,s,k}}$ at MBS 0, respectively.
\par
We assume that all computation tasks of each IMD are executed sequentially to satisfy practical implementations. However, local execution and computation offloading can be performed for any task in a parallel manner. Therefore, the total time ${\tau}_{i}$ used for completing all task of IMD $i$ can be given by
\begin{equation}\label{eq12}
	{\tau}_{i}=\sum\nolimits_{k\in \mathcal{K}}{\max \big( \sum\nolimits_{s\in \mathcal{S}}{{{x}_{i,s}}{\tau}_{_{i,s,k}}^{LOC}},\sum\nolimits_{s\in {\mathcal{S}}}{{{x}_{i,s}}{\tau}_{_{i,s,k}}^{BS}} \big)},
\end{equation}
\par
Then, the total energy consumed by all IMDs can be given by
\begin{equation}\label{eq13}
\begin{split}
  &\epsilon=\sum\nolimits_{i\in \mathcal{U}}{\sum\nolimits_{k\in \mathcal{K}}{\sum\nolimits_{s\in \mathcal{S}}{\varsigma {{x}_{i,s}}\left( {{d}_{i,k}}-{{{\bar{d}}}_{i,s,k}} \right){{c}_{i,k}}f_{i}^{2}}}} \\
 & \ \ \ \ +\sum\nolimits_{i\in \mathcal{U}}{\sum\nolimits_{k\in \mathcal{K}}{\sum\nolimits_{s\in \mathcal{S}}{\sum\nolimits_{l\in \mathcal{L}}{{{x}_{i,s}}{{y}_{i,k,l}}{{{\tilde{\theta }}}_{l}}{{{\bar{d}}}_{i,s,k}}}}}} \\
 & \ \ \ +\sum\nolimits_{i\in \mathcal{U}}{\sum\nolimits_{k\in \mathcal{K}}{\sum\nolimits_{s\in \mathcal{S}}{\sum\nolimits_{n\in \mathcal{N}}{{{x}_{i,s}}{{z}_{i,n}}{{p}_{i}}{{{{\bar{d}}}_{i,s,k}}}/{{{R}_{i,s,n}}}}}}}, \\
\end{split}
\end{equation}
where $\varsigma$ is the energy coefficient of chip architecture; the three items on the right side of \eqref{eq13} are total computing, encrypting and uploading energy consumptions of IMDs, respectively.
\section{PROBLEM FORMULATION and SOLUTION}\label{sec 3}
\subsection{Problem Formulation}
To achieve the goal of green and secure computation offloading in ultra-dense multi-task IoT Networks, we try to minimize the total energy consumption of all IMDs under joint BS clustering, OFDMA and NOMA, proportional allocation of computational resources, and the constraints of latency and security costs, jointly optimizing the device association, channel selection, security service assignment, power control and computation computing resources. Mathematically, it can be formulated as
\begin{equation}\label{eq14}
	\begin{aligned}
		&\underset{\mathbf{X},\mathbf{Y},\mathbf{Z},\mathbf{p},\mathbf{\bar{D}},\mathbf{\hat{D}}}{\min}\,  \epsilon\big(\mathbf{X},\mathbf{Y},\mathbf{Z},\mathbf{p},\mathbf{\bar{D}},\mathbf{\hat{D}}  \big)\\
&\text{s.t. }{{C}_{1}}:{\tau}_{i}\le {\tau}_{i}^{\max },\forall i\in \mathcal{U}, \\
		& {{C}_{2}}:{{\psi}_{i}}\le {\psi}_{i}^{\max },\forall i\in \mathcal{U}, \\
		& {{C}_{3}}:\sum\nolimits_{s\in \mathcal{S}}{{{x}_{i,s}}}=1,\forall i\in \mathcal{U}, \\
		& {{C}_{4}}:\sum\nolimits_{l\in \mathcal{L}}{{{y}_{i,k,l}}}=1,\forall i\in \mathcal{U},\forall k\in \mathcal{K}, \\
        & {{C}_{5}}:\sum\nolimits_{i\in \mathcal{U}}{{{z}_{i,n}}}=1,\forall n\in \mathcal{N},\\
		& {{C}_{6}}:\vartheta \le {{p}_{i}}\le p_{i}^{\max },\forall i\in \mathcal{U}, \\
		& {{C}_{7}}:{{x}_{i,s}}\in \left\{ 0,1 \right\},\forall i\in \mathcal{U},s\in \mathcal{S}, \\
		& {{C}_{8}}:{{y}_{i,k,l}}\in \left\{ 0,1 \right\},\forall i\in \mathcal{U},\forall k\in \mathcal{K},l\in \mathcal{L}, \\
		& {{C}_{9}}:{{z}_{i,n}}\in \left\{ 0,1 \right\},\forall i\in \mathcal{U},\forall n\in \mathcal{N}, \\
		& {{C}_{10}}:\theta \le {{{\hat{d}}}_{i,s,k}}\le {{{\bar{d}}}_{i,s,k}}\le {{d}_{i,k}},\forall i\in \mathcal{U},s\in \mathcal{S},k\in \mathcal{K}, \\
	\end{aligned}	
\end{equation}
where $\mathbf{X}=\left\{{{x}_{i,s}},\forall i\in \mathcal{U},\forall s\in \mathcal{S} \right\}$, $\mathbf{Y}=\{{{y}_{i,k,l}},\forall i\in \mathcal{U},\forall k\in \mathcal{K},\forall l\in \mathcal{L}\}$, $\mathbf{Z}=\left\{{{z}_{i,n}},\forall i\in \mathcal{U},\forall n\in \mathcal{N} \right\}$, $\mathbf{p}=\left\{ {{p}_{i}},\forall i\in \mathcal{U} \right\}$, $\mathbf{\bar{D}}=\big\{ {{{\bar{d}}}_{i,s,k}},\forall i\in \mathcal{U},\forall s\in \mathcal{S},\forall k\in \mathcal{K} \big\}$, $\mathbf{\hat{D}}=\big\{ {{{\hat{d}}}_{i,s,k}},\forall i\in \mathcal{U},\forall s\in \mathcal{S},\forall k\in \mathcal{K} \big\}$; $\vartheta $ takes a small enough value to avoid zero division, e.g., ${{10}^{\text{-}20}}$; ${{C}_{1}}$ indicates that the task execution time of IMD $i$ cannot exceed its deadline ${\tau}_{i}^{\max }$; ${{C}_{2}}$ means that total security breach cost of IMD $i$ cannot exceed its maximum acceptable cost ${\psi}_{i}^{\max }$ ; ${{C}_{3}}$ and ${{C}_{7}}$ indicate that an IMD can just be associated with only one BS; ${{C}_{4}}$ and ${{C}_{8}}$ indicate that the task $k$ of IMD $i$ can just select only one cryptographic algorithm; ${{C}_{5}}$ and ${{C}_{9}}$ indicate that an IMD can just select only one subchannel; ${{C}_{6}}$ gives the lower bound ($\vartheta$) and upper bound ($p_{i}^{\max }$) of the transmission power of IMD $i$; ${{C}_{10}}$ means that the offloaded parts ${{\bar{d}}_{i,s,k}}$ and ${{\hat{d}}_{i,s,k}}$ are greater than or equal to $\vartheta$, but less than or equal to the data size ${{{d}}_{i,k}}$ of task $k$ of IMD $i$. Meanwhile, ${{\hat{d}}_{i,s,k}}$ must be less than or equal to ${{\bar{d}}_{i,s,k}}$.
\subsection{Algorithm Design}
As revealed in \cite{QVPham2020Apr}, WOA is a gradient-free method and can relax the computations of gradients. In addition, it is insensitive to the initial feasible solutions, which may affect the convergence and performance of other traditional methods greatly. Moreover, WOA has been equipped with adaptive mechanisms that balance its explorative and exploitative behaviors appropriately, which can increase the probability of avoiding locally optimal solutions. At last, since WOA is flexible and easy to be implemented, it is applicable to common optimization problems rather than particular ones. So far, WOA has been regarded as a promising solution to the optimization problem in wireless and communication networks. In view of this, we develop IWOA to solve the formulated problem \eqref{eq14} by improving WOA in \cite{QVPham2020Apr}, which consists of encircling prey, bubble-net attacking (exploitation phase) and prey search (exploration phase). To utilize IWOA to solve \eqref{eq14}, we need to encode whales, define their fitness function, and initialize their values. Then, the procedures of encircling prey, bubble-net attacking and search for prey are given in detail.
\par
\textbf{\textit{1) Encode whale}}
\par
The optimization parameters $\mathbf{X}$, $\mathbf{Y}$, $\mathbf{Z}$, $\mathbf{p}$, $\mathbf{\bar{D}}$, $\mathbf{\hat{D}}$ of problem \eqref{eq14} are encoded as ${{\mathbf{B}}_{m}}$, ${{\mathbf{O}}_{m}}$, ${{\mathbf{E}}_{m}}$,  ${{\mathbf{Q}}_{m}}$, ${{\mathbf{G}}_{m}}$, ${{\mathbf{H}}_{m}}$ respectively, where ${{\mathbf{B}}_{m}}=\left\{ {{b}_{m,i}},i\in \mathcal{U} \right\}$, and ${{b}_{m,i}}$ is the BS index selected by IMD $i$ in the individual (whale) $m$; ${{{\mathbf O}}_{m}}=\left\{ {{o}_{m,i}},i\in \bar{\mathcal{U}} \right\}$, $\bar{\mathcal{U}}=\{1,2,\cdots,K,K+1,\cdots,2K,\cdots,UK\}$, and ${{o}_{m,i}}$ is the index of cryptographic algorithm selected by virtual IMD $i$ in the individual $m$; ${{\mathbf{E}}_{m}}=\left\{ {{e}_{m,i}},i\in \mathcal{U} \right\}$, and ${{e}_{m,i}}$ is the channel index selected by IMD $i$ in the individual $m$; ${{\mathbf{Q}}_{m}}=\left\{ {{q}_{m,i}},i\in \mathcal{U} \right\}$, and ${{q}_{m,i}}$ is the transmission power of IMD $i$ in the individual $m$; ${{\mathbf{G}}_{m}}=\left\{ {{g}_{m,i}},i\in \bar{\mathcal{U}} \right\}$, and ${{g}_{m,i}}$ is the amount of data offloaded from IMD $i$ to its associated SBS in the individual $m$; ${{\mathbf{H}}_{m}}=\left\{ {{h}_{m,i}},i\in \bar{\mathcal{U}} \right\}$, and ${{h}_{m,i}}$ is the amount of data offloaded from IMD $i$ or its associated SBS to nearby MBS in the individual $m$. Significantly, ${\mathcal{M}}=\{1,2,\cdots,M\}$ represents the population consisting of $M$ individuals (whales).
\par
\begin{figure}[!t]
	\centering
	\centerline{\includegraphics[width=3in]{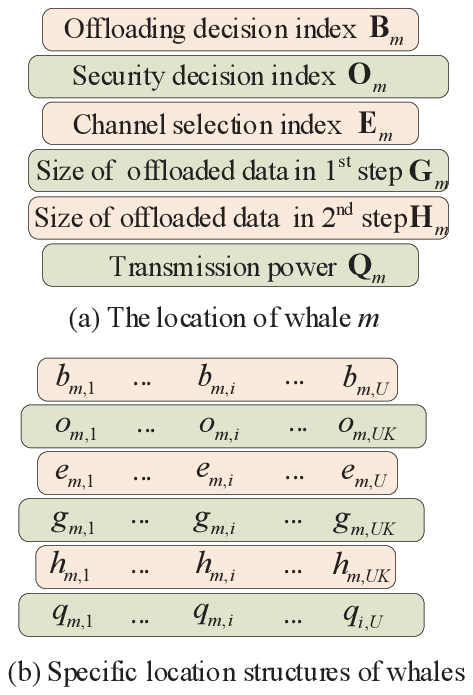}}
	\caption{Encoding structures of whales.}
	\label{fig2}
\end{figure}

The coding and structure of individuals are shown in Fig.\ref{fig2}.
\par
\textbf{\textit{2) Fitness function}}
\par
To assess the fitness of individuals (whales), fitness functions need to be designed properly. Seen from \eqref{eq14}, we can easily observe that constraints ${{C}_{1}}$ and ${{C}_{2}}$ are in nonlinear, mixed-integer and coupling forms, and hard to be met in whales' actions. In view of this, they are introduced into the fitness function as penalty terms, which can be explicitly used to prevent individuals from falling into the infeasible region. In this way, the established population can always find a feasible optimal solution.
\par
To minimize the energy consumed by all IMDs under the constraints ${{C}_{1}}$ and ${{C}_{2}}$, the fitness function of individual $m$ can be defined as
\begin{equation}\label{eq15}
\begin{aligned}
&F({\mathbf{B}_{m}},{\mathbf{O}_{m}},{\mathbf{E}_{m}},{{\mathbf{Q}}_{m}},{{\mathbf{G}}_{m}},{{\mathbf{H}}_{m}})\\ &\ \ =-{\epsilon}({\mathbf{B}_{m}},{\mathbf{O}_{m}},{\mathbf{E}_{m}},{{\mathbf{Q}}_{m}},{{\mathbf{G}}_{m}},{{\mathbf{H}}_{m}})\\
&\ \ \ \ \ -\sum\nolimits_{i\in \mathcal{U}}{{{\alpha }_{i}}\max\left( 0,{\tau}_{i}-{\tau}_{i}^{\max } \right)}\\
&\ \ \ \ \ -\sum\nolimits_{i\in \mathcal{U}}{{{\beta }_{i}}\max\left( 0,{{\psi }_{i}}-{\psi}_{i}^{\max} \right)}, \\
\end{aligned}
\end{equation}
where ${{\alpha }_{i}}$ and ${{\beta }_{i}}$ are the penalty factors of IMD $i$.
\par
\textbf{\textit{3) Population initialization}}
\par
In order to meet the constraints ${{C}_{3}}$-${{C}_{10}}$, initial population can be generated use the following rules. Specifically, any individual $m$ can be initialized into
\begin{equation}\label{eq16}
\left\{ \begin{aligned}
& b_{m,i}^{0}=\text{randi}(\mathcal{S}),\forall i\in \mathcal{U}, \\
& o_{m,i}^{0}=\text{randi}(\mathcal{L}),\forall i\in \bar{\mathcal{U}}, \\
& e_{m,i}^{0}=\text{randi}(\mathcal{N}),\forall i\in \bar{\mathcal{U}}, \\
& q_{m,i}^{0}=\text{rand}\left( p_{i}^{\max } \right),\forall i\in \mathcal{U}, \\
& g_{m,i}^{0}=\text{rand}\left( {{d}_{u,k}} \right),\forall i\in \bar{\mathcal{U}}, \\
& h_{m,i}^{0}=\text{rand}\left( g_{m,i}^{0} \right),\forall i\in \bar{\mathcal{U}}, \\
& [u,k]=\text{ind2sub}([U K],i),\forall i\in \bar{\mathcal{U}}, \\
\end{aligned} \right.
\end{equation}
where $[u,k]=\text{ind2sub}([U K],i)$ returns the row subscript $m$ and column subscript $k$ of $U\times K$ matrix corresponding to the linear index $i$; $\text{randi}(\mathcal{Z})$ outputs an element from the set $\mathcal{Z}$ randomly, and $\text{rand}(\gamma)$ generates a random number between 0 and $\gamma$.
\par
\textbf{\textit{4) Encircle prey}}
\par
Humpback whales can recognize the locations of prey and then encircle them completely. Therefore, all whales are agents that search for prey. In conventional WOA \cite{QVPham2020Apr}, the current best agent is assumed to be the target prey, and all whales update their positions towards it during iteration. To ensure global convergence of WOA, we replace the current best agent with the historically best agent. The former refers to the individual (whale) that owns the highest fitness function value among all individuals in the current iteration, but the latter refers to the one that has the highest fitness function value among all individuals in the previous and current iterations. Mathematically, the behavior of encircling prey $\bar{m}$ of individual (whale) $m$ can be formulated as
\begin{equation}\label{eq17}
{{b}_{m,i}}=\text{round}\left( {\kappa}_{1}{{b}_{\bar{m},i}}-{{\kappa}_{2}}\left| {{\kappa}_{3}}{{b}_{\bar{m},i}}-{{b}_{m,i}} \right| \right),\forall i\in \mathcal{U},	
\end{equation}
\begin{equation}\label{eq18}
{{o}_{m,i}}=\text{round}\left( {\kappa}_{1}{{o}_{\bar{m},i}}-{{\kappa}_{2}}\left| {{\kappa}_{3}}{{o}_{\bar{m},i}}-{{o}_{m,i}} \right| \right),\forall i\in \bar{\mathcal{U}},
\end{equation}
\begin{equation}\label{eq19}
{{e}_{m,i}}=\text{round}\left( {\kappa}_{1}{{e}_{\bar{m},i}}-{{\kappa}_{2}}\left| {{\kappa}_{3}}{{e}_{\bar{m},i}}-{{e}_{m,i}} \right| \right),\forall i\in \mathcal{U},
\end{equation}
\begin{equation}\label{eq20}
{{q}_{m,i}}={\kappa}_{1}{{q}_{\bar{m},i}}-{{\kappa}_{2}}\left| {{\kappa}_{3}}{{q}_{\bar{m},i}}-{{q}_{m,i}} \right|,\forall i\in \mathcal{U},
\end{equation}
\begin{equation}\label{eq21}
{{g}_{m,i}}={\kappa}_{1}{{g}_{\bar{m},i}}-{{\kappa}_{2}}\left| {{\kappa}_{3}}{{g}_{\bar{m},i}}-{{g}_{m,i}} \right|,\forall i\in \bar{\mathcal{U}},
\end{equation}
\begin{equation}\label{eq22}
{{h}_{m,i}}={\kappa}_{1}{{h}_{\bar{m},i}}-{{\kappa}_{2}}\left| {{\kappa}_{3}}{{h}_{\bar{m},i}}-{{h}_{m,i}} \right|,\forall i\in \bar{\mathcal{U}},
\end{equation}
where $\text{round}(\gamma )$ represents a rounding operation on $\gamma$; $|\gamma |$ is the absolute value of $\gamma$; $\bar{m}$ is the index of historically best agent (individual);
\begin{equation}\label{eq23}
\left\{ \begin{aligned}
& {{\kappa}_{1}}=\sin \left( {t\pi }/{2T}+\pi  \right)+1, \\
& {{\kappa}_{2}}=2\left( 2{{r}_{1}}-1 \right)\left( 1-\sin \left( {t\pi}/{2T}\right)\right), \\
& {{\kappa}_{3}}=2{{r}_{2}}, \\
\end{aligned} \right.
\end{equation}
${r}_{1}$ and ${r}_{2}$ are random numbers between 0 and 1; $t$ is iteration index; $T$ is the number of iterations.
\par
Inspired by the efforts in \cite{ZGuo2017Sep,ZWu2020dec}, adaptive nonlinear weights ${\kappa}_{1}$ and ${\kappa}_{2}$ are introduced for updating the positions of individuals (whales) in \eqref{eq17}-\eqref{eq22}. In addition, these weights are also used for bubble-net attacks and searching for prey. It is easy to find that such weights can balance exploitation and exploration well.
\par
\textbf{\textit{5) Bubble-net attacking }}
\par
Bubble-net attacking of humpback whales involves shrinking encircling and spiral movement simultaneously, which are performed in equal probability. By performing these actions, the new position of any agent will be located between its current position and the position of the historically best agent. It means that a local optimum of problem \eqref{eq14} can be found using bubble-net attacking. To mimic the helix-shaped movement of whales, the spiral equation between the positions of prey $\bar{m}$ of any individual (whale) $m$ can be given by
\begin{equation}\label{eq24}
{{b}_{m,i}}=\text{round}\left( {\kappa}_{1}{{b}_{\bar{m},i}}+{{\kappa}_{4}}\left| {{\kappa}_{3}}{{b}_{\bar{m},i}}-{{b}_{m,i}} \right|\right),\forall i\in \mathcal{U},	
\end{equation}
\begin{equation}\label{eq25}
{{o}_{m,i}}=\text{round}\left( {\kappa}_{1}{{o}_{\bar{m},i}}+{{\kappa}_{4}}\left| {{\kappa}_{3}}{{o}_{\bar{m},i}}-{{o}_{m,i}} \right|\right),\forall i\in \bar{\mathcal{U}},
\end{equation}
\begin{equation}\label{eq26}
{{e}_{m,i}}=\text{round}\left( {\kappa}_{1}{{e}_{\bar{m},i}}+{{\kappa}_{4}}\left| {{\kappa}_{3}}{{e}_{\bar{m},i}}-{{e}_{m,i}} \right|\right),\forall i\in \mathcal{U},
\end{equation}
\begin{equation}\label{eq27}
{{q}_{m,i}}={\kappa}_{1}{{q}_{\bar{m},i}}+{{\kappa}_{4}}\left| {{\kappa}_{3}}{{q}_{\bar{m},i}}-{{q}_{m,i}} \right|,\forall i\in \mathcal{U},
\end{equation}
\begin{equation}\label{eq28}
{{g}_{m,i}}={\kappa}_{1}{{g}_{\bar{m},i}}+{{\kappa}_{4}}\left| {{\kappa}_{3}}{{g}_{\bar{m},i}}-{{g}_{m,i}} \right|,\forall i\in \bar{\mathcal{U}},
\end{equation}
\begin{equation}\label{eq29}
{{h}_{m,i}}={\kappa}_{1}{{h}_{\bar{m},i}}+{{\kappa}_{4}}\left| {{\kappa}_{3}}{{h}_{\bar{m},i}}-{{h}_{m,i}} \right|,\forall i\in \bar{\mathcal{U}},
\end{equation}
where ${{\kappa}_{4}}$ is used for adaptively adjusting the spiral amplitude and avoiding falling into local optimum \cite{LiuL2020}, and it can be given by
\begin{equation}\label{eq30}
\left\{ \begin{aligned}
&{{\kappa}_{4}}=\exp \left( {{a}_{3}}+5\cos \left( \pi \left( 1-{t}/{T} \right) \right) \right)\cos \left( 2{{a}_{3}}\pi  \right),\\
&{{a}_{3}}=\left( -2-{t}/{T} \right)*{{r}_{3}}+1, \\
\end{aligned} \right.
\end{equation}
where ${r}_{3}$ is a random number between 0 and 1.
\par
Besides spiral movement, any whale also needs to perform shrinking encircling action during the bubble-net attacking phase, which can be formulated as \eqref{eq17}-\eqref{eq22}.
\par
\textbf{\textit{6) Search for prey}}
\par
In the prey search of conventional WOA, whales are forced to move toward a random whale. Through such an operation, the search space of this algorithm can be extended. However, its global search capability may greatly rely on the selection of the random whale, and it may be easy to fail into local optimum. To tackle this issue, Cauchy's inverse cumulative distribution function may be used for the mutation operations of whales since its long tail \cite{LiuL2020}. Inspired by this point, Cauchy's inverse cumulative distribution is used for formulating the prey search of whales. Mathematically, the behavior of prey search of any individual (whale) $m$ can be formulated as
\begin{equation}\label{eq31}
{{b}_{m,i}}=\operatorname{round}\left( {{b}_{m,i}}+{{\kappa}_{2}}\tan \left( \pi \left( {{r}_{1}}-0.5 \right) \right) \right),\forall i\in \mathcal{U},
\end{equation}
\begin{equation}\label{eq32}
{{o}_{m,i}}=\operatorname{round}\left( {{o}_{m,i}}+{{\kappa}_{2}}\tan \left( \pi \left( {{r}_{1}}-0.5 \right) \right) \right),\forall i\in \bar{\mathcal{U}},
\end{equation}
\begin{equation}\label{eq33}
{{e}_{m,i}}=\operatorname{round}\left( {{e}_{m,i}}+{{\kappa}_{2}}\tan \left( \pi \left( {{r}_{1}}-0.5 \right) \right) \right),\forall i\in \mathcal{U},
\end{equation}
\begin{equation}\label{eq34}
{{q}_{m,i}}={{q}_{m,i}}+{{\kappa}_{2}}\tan \left( \pi \left( {{r}_{1}}-0.5 \right) \right),\forall i\in \mathcal{U},
\end{equation}
\begin{equation}\label{eq35}
{{g}_{m,i}}={{g}_{m,i}}+{{\kappa}_{2}}\tan \left( \pi \left( {{r}_{1}}-0.5 \right) \right),\forall i\in \bar{\mathcal{U}},
\end{equation}
\begin{equation}\label{eq36}
{{h}_{m,i}}={{h}_{m,i}}+{{\kappa}_{2}}\tan \left( \pi \left( {{r}_{1}}-0.5 \right) \right),\forall i\in \bar{\mathcal{U}},
\end{equation}
It is noteworthy that weight ${{\kappa}_{2}}$ is used for adaptively adjusting the magnitude of mutation.
\par
\textbf{\textit{6)  Search for prey in the nearby area of historically best agent}}
\par
In order to improve the convergence rate, avoid premature convergence and thus achieve a better solution, we further force whales (individuals) to search for prey in the nearby area of the historically best agent once more. Mathematically, in the nearby area of the historically best agent $\bar{m}$, new positions of any individual (whale) $m$ can be generated by
\begin{equation}\label{eq37}
{\bar{b}_{m,i}}=\text{round}\left( {{b}_{\bar{m},i}}\left( 1+0.5{{r}_{4}} \right) \right),\forall i\in \mathcal{U},
\end{equation}
\begin{equation}\label{eq38}
{\bar{o}_{m,i}}=\text{round}\left( {{o}_{\bar{m},i}}\left( 1+0.5{{r}_{4}} \right) \right),\forall i\in \bar{\mathcal{U}},
\end{equation}
\begin{equation}\label{eq39}
{\bar{e}_{m,i}}=\text{round}\left( {{e}_{\bar{m},i}}\left( 1+0.5{{r}_{4}} \right) \right),\forall i\in \mathcal{U},
\end{equation}
\begin{equation}\label{eq40}
{\bar{q}_{m,i}}={{q}_{\bar{m},i}}\left( 1+0.5{{r}_{4}} \right),\forall i\in \mathcal{U},
\end{equation}
\begin{equation}\label{eq41}
{\bar{g}_{m,i}}={{g}_{\bar{m},i}}\left( 1+0.5{{r}_{4}} \right),\forall i\in \bar{\mathcal{U}},
\end{equation}
\begin{equation}\label{eq42}
{\bar{h}_{m,i}}={{h}_{\bar{m},i}}\left( 1+0.5{{r}_{4}} \right),\forall i\in \bar{\mathcal{U}},
\end{equation}
where ${r}_{4}$ is a random number between 0 and 1.
\par
As for the newly generated positions using \eqref{eq37}-\eqref{eq42}, we decide whether or not to save them in a greedy approach \cite{LiuL2020}. Specifically, the original positions of whales should be replaced with them when $F({{\mathbf{B}}_{m}},{{\mathbf{O}}_{m}},{{\mathbf{E}}_{m}},{{\mathbf{Q}}_{m}},{{\mathbf{G}}_{m}},{{\mathbf{H}}_{m}})\le F({{\mathbf{\bar{B}}}_{m}},{{\mathbf{\bar{O}}}_{m}},{{\mathbf{\bar{E}}}_{m}},{{\mathbf{\bar{Q}}}_{m}},{{\mathbf{\bar{G}}}_{m}},{{\mathbf{\bar{H}}}_{m}})$, where ${\bar{\mathbf{B}}_{m}}=\left\{ {\bar{b}_{m,i}},i\in \mathcal{U} \right\}$, ${{\bar{\mathbf O}}_{m}}=\left\{ {\bar{o}_{m,i}},i\in \bar{\mathcal{U}} \right\}$, ${\bar{\mathbf{E}}_{m}}=\left\{ {\bar{e}_{m,i}},i\in \mathcal{U} \right\}$, ${\bar{\mathbf{Q}}_{m}}=\left\{ {\bar{q}_{m,i}},i\in \mathcal{U} \right\}$, ${\bar{\mathbf{G}}_{m}}=\left\{ {\bar{g}_{m,i}},i\in \bar{\mathcal{U}} \right\}$ and ${\bar{\mathbf{H}}_{m}}=\left\{ {\bar{h}_{m,i}},i\in \bar{\mathcal{U}} \right\}$. Otherwise, the original positions of whales should not be changed.
\par
Until now, the whole procedure used for IWOA can be summarized as Algorithm 1.
\begin{table}[]
	\centering
	\begin{tabular}{ll}
		\toprule[1pt]
		\textbf{Algorithm 1: Improved WOA (IWOA)} \\ \midrule[0.5pt]
        1: \textbf{Input:} Number $T$ of iterations. \\
        2: \textbf{Output:} $\mathbf{B}$, $\mathbf{O}$, $\mathbf{E}$, $\mathbf{Q}$, $\mathbf{G}$ and $\mathbf{H}$ at $t$-th iteration. \\
		3: \textbf{Initialization:}\\
        4:\ \ \ \ Initialize iteration index: $t=1$.\\
		5:\ \ \ \ Initialize the population consisting of $M$ agents using \eqref{eq16}.\\
		6:\ \ \ \ Calculate the fitness values of all agents using \eqref{eq15}.\\
		7:\ \ \ \ Find the historically best agent among all agents.\\
		8: \textbf{While $t<=T$} \textbf{do}\\
		9:\ \ \ \ Update ${{\kappa}_{1}}$, ${{\kappa}_{2}}$, ${{\kappa}_{3}}$ and ${{\kappa}_{4}}$ using \eqref{eq23} and \eqref{eq30}.\\
		10:\ \ \ \ Generate the probability ${r}_{5}$ randomly.\\
		11:\ \ \ \ \textbf{If } ${r}_{5}<0.5$ holds, then \\
		12:\ \ \ \ \ \ \textbf{If} $|{{\kappa}_{2}}|\ge 1$ holds, then \\
		13:\ \ \ \ \ \ \ \ All agents search prey using \eqref{eq31}-\eqref{eq36}.\\
		14:\ \ \ \ \ \ \textbf{Else} \\
		15:\ \ \ \ \ \ \ \ All agents encircle prey using \eqref{eq17}-\eqref{eq22}.\\
		16:\ \ \ \ \ \ \textbf{EndIf} \\
		17:\ \ \ \ \textbf{Else} \\
		18:\ \ \ \ \ \ All agents perform bubble-net attacks using \eqref{eq24}-\eqref{eq29}. \\
		19:\ \ \ \ \textbf{EndIf} \\
		20:\ \ \ \ Calculate fitness value ${{\chi }_{m}}=F({{\mathbf{B}}_{m}},{{\mathbf{O}}_{m}},{{\mathbf{E}}_{m}},{{\mathbf{Q}}_{m}},{{\mathbf{G}}_{m}},{{\mathbf{H}}_{m}})$\\
		21:\ \ \ \ \ \ of any agent $m$ using \eqref{eq15}.\\
		22:\ \ \ \ Find the current best agent, and replace historically best agent\\
		23:\ \ \ \ \ \ with it if its fitness value is higher than historically best agent.\\
		24:\ \ \ \ Any agent searches for prey in the nearby area of historically best\\
 		25:\ \ \ \ \ \ agent, and generates new position $\{{{\mathbf{\bar{B}}}_{m}},{{\mathbf{\bar{O}}}_{m}},{{\mathbf{\bar{E}}}_{m}},{{\mathbf{\bar{Q}}}_{m}},{{\mathbf{\bar{G}}}_{m}},$\\
 		26:\ \ \ \ \ \ ${{\mathbf{\bar{H}}}_{m}}\}$ using \eqref{eq37}-\eqref{eq42}.\\
 		27:\ \ \ \ Calculate fitness value ${\bar{\chi }_{m}}=F({{\mathbf{\bar{B}}}_{m}},{{\mathbf{\bar{O}}}_{m}},{{\mathbf{\bar{E}}}_{m}},{{\mathbf{\bar{Q}}}_{m}},{{\mathbf{\bar{G}}}_{m}},{{\mathbf{\bar{H}}}_{m}})$\\
 		28:\ \ \ \ \ \ of any agent $m$ using \eqref{eq15}.\\
 		29:\ \ \ \ \textbf{If } ${\bar{\chi }_{m}}>{{\chi }_{m}}$ holds, then \\
		30:\ \ \ \ \ \ $\{{{\mathbf{{B}}}_{m}},{{\mathbf{{O}}}_{m}},{{\mathbf{{E}}}_{m}},{{\mathbf{{Q}}}_{m}},{{\mathbf{{G}}}_{m}},{{\mathbf{{H}}}_{m}}\}$ is replaced with $\{{{\mathbf{\bar{B}}}_{m}},{{\mathbf{\bar{O}}}_{m}},{{\mathbf{\bar{E}}}_{m}},$\\
 		31:\ \ \ \ \ \ \ \	${{\mathbf{\bar{Q}}}_{m}},{{\mathbf{\bar{G}}}_{m}},{{\mathbf{\bar{H}}}_{m}}\}$ \\
  		32:\ \ \ \ \textbf{EndIf}\\
		33:\ \ \ Update the iteration index: ${t}={t}+1$.\\
		34: \textbf{EndWhile}\\ \bottomrule[0.5pt]
	\end{tabular}
	\label{alg1}
\end{table}

\section{ALGORITHM ANALYSIS}\label{sec 5}
In this section, the convergence, computational complexity, and parallel implementation of IWOA will be analyzed in detail.
\subsection{Convergence Analysis}
The convergence of IWOA can be established as follows.
\par
\noindent
\textbf{\textit{Theorem 1:}} IWOA converges to global optimum solution after a large number of iterations.
\par
\textit{Proof:} In Algorithm 1 (IWOA), all whales perform the encircling prey, bubble-net attacking (exploitation phase) and prey search (exploration phase) in Steps 9-19. When the iteration index $t$ gradually approaches $T$, ${\kappa}_{2}$ is closer and closer to zero. Evidently, when $t=T$, all whales don't search for prey using \eqref{eq31}-\eqref{eq36}, they encircle prey and perform bubble-net attacks in equal probability. In other words, IWOA only contains two operations consisting of shrinking encirclement and spiral update at this time. Such operations are performed by whales in equal probability. Even if a common whale falls into a local optimum solution during the spiral update, it may jump out of such a solution when the shrinking operation is done.
\par
It is noteworthy that the historically best agent (whale/individual) always remains in the population of IWOA. In addition, all agents led by this agent perform the operations of shrinking encirclement and spiral update. It means that these two operations force all agents to move toward the historically best agent. Evidently, when the number of iterations of IWOA tends to infinity, it can finally converge to the global optimum solution.
\par
In Steps 20-32 of IWOA, all agents are forced to search for prey in the nearby area of the historically best agent. Such an operation can refine the solutions found by agents, and improve the historically best agent. It is evident that Steps 20-32 of IWOA can speed up the global convergence of this algorithm.
\par
In general, IWOA converges to a global optimum solution after a large number of iterations.
\ding{113}
\subsection{Complexity Analysis}
The computational complexity of IWOA is analyzed as follows.\\
\par
\noindent
\textbf{\textit{Proposition 1: }}The computational complexity of IWOA is $\mathcal{O}\left(\max\{TMUK, TMNU^2 \right)$ after $T$ iterations in the worst scenario that all IMDs share each channel.
\par
In Steps 6-7, the computational complexity is mainly dependent on the calculation of the fitness values of all agents. These fitness values are tightly related to energy consumption, delay and security breach cost of all IMDs. In fact, it is easy to find that the computational complexities of energy consumption and delay mainly come from the calculations of data rates and computing capabilities. To calculate the data rates and computing capabilities, we first convert $\bar{\mathbf{B}}_{m}$ and $\bar{\mathbf{E}}_{m}$ into $\mathbf{a}=\{{a}_{i},\forall i\in \mathcal{U}\}$ and $\mathbf{b}=\{{b}_{i},\forall i\in \mathcal{U}\}$ for any individual $m$, respectively. In addition, $\bar{\mathbf{O}}_{m}$ is converted into the indices of cryptographic algorithms for any individual $m$. Through these conversions, the calculations of data rates and computing capabilities can be greatly reduced since we just need to consider the utilized BSs, channels and cryptographic algorithms for any IMD.
\par
In \eqref{eq1}, $\sum\nolimits_{u\in {{\mathcal{Q}}_{i,s,n}}}{{{p}_{u}}{{\hbar}_{u,s}}}$ can be calculated before calculating ${{R}_{i,s,n}}$. Similarly, $\sum\nolimits_{u\in \mathcal{U}}{{{x}_{u,0}}}$ can be calculated before calculating ${{R}_{i,0,n}}$. Consequently, under the given ${a}_{i}$ and ${b}_{i}$,  the calculation of ${{R}_{i,s,n}}$ may have a complexity of $\mathcal{O}\left(NU^2\right)$ in the worst scenario that all IMDs share each channel, and the one of ${{R}_{i,0,n}}$ may have a complexity of $\mathcal{O}\left(NU\right)$. In \eqref{eq8}, $\sum\nolimits_{l\in \mathcal{L}}{{{y}_{i,k,l}}\bar{\Gamma}_{i,s,k,l}}$ and $ \sum\nolimits_{u\in \mathcal{U}}{\sum\nolimits_{j\in \mathcal{K}}{{{x}_{u,s}}\left({\Gamma}_{u,s,j}+ \sum\nolimits_{l\in \mathcal{L}}{{{y}_{u,j,l}}\bar{\Gamma}_{u,s,j,l}} \right)}} $ can be calculated before calculating ${{{\bar{f}}}_{i,s,k}}$ for SBS $s$. In \eqref{eq10}, $\sum\nolimits_{s\in \bar{\mathcal{S}}}{{{x}_{i,s}}{{\Upsilon }_{u,s,j}}}$ and $\sum\nolimits_{u\in \mathcal{U}}{\sum\nolimits_{j\in \mathcal{K}}{\left( \sum\nolimits_{s\in \bar{\mathcal{S}}}{{{x}_{u,s}}{{\Upsilon }_{u,s,j}}}+{{x}_{u,0}}{{{\bar{\Upsilon }}}_{u,0,j}} \right)}}$ can be calculated before calculating ${{\bar{f}}_{i,0,k}}$ for MBS 0. Consequently, under the given ${a}_{i}$ and ${b}_{i}$, the computational complexity of ${{\bar{f}}_{i,s,k}}$ is $\mathcal{O}\left(USK \right)$ for any BS $s$.
\par
Based on the above-mentioned analyses, under the given $\mathbf{a}$, $\mathbf{b}$ and indices of cryptographic algorithms, the computational complexity of delay is  $\mathcal{O}\left(\max\{UK, NU^2 \right)$ for all IMDs in the worst scenario, and the one of total energy consumption $\epsilon$ is still $\mathcal{O}\left(\max\{UK, NU^2 \right)$ in the worst scenario. In addition, it is easy to find that the computational complexity of security breach cost is $\mathcal{O}\left(UK\right)$ for all IMDs under the given indices of cryptographic algorithms. In general, Steps 6-7 have a computational complexity of $\mathcal{O}\left(\max\{MUK, MNU^2 \right)$ in the worst scenario.
\par
Evidently, the computational complexity of Steps 9-19 is $\mathcal{O}\left(MUK \right)$. The one of Steps 20-33 mainly comes from the calculation of fitness values of all agents, which is $\mathcal{O}\left(\max\{MUK, MNU^2 \right)$ in the worst scenario. After $T$ iterations, the one of IWOA is $\mathcal{O}\left(\max\{TMUK, TMNU^2 \right)$ in the worst scenario.
\ding{113}
\subsection{Parallel implementation}
As revealed in the previous section, the computational complexity of IWOA mainly comes from the calculations of the fitness values of all agents. Such calculations will lead to relatively high computational complexity if the number of agents is too large. In order to reduce computational complexity and improve the efficiency of designed algorithm, all agents should calculate their fitness values in a parallel manner, which has been widely advocated in reality. Certainly, any one of three operations consists of encircling prey, bubble-net attacking and prey search can be also performed by all agents in parallel.

\section{NUMERICAL RESULTS}\label{sec 6}
Without loss of generality, IMDs and ultra-dense SBSs are randomly deployed into a macrocell, where the number of SBSs is greater than or equal to the number of IMDs. At the same time, we consider $\boldsymbol{\bar{\theta}}=[100, 200, 250, 300, 350, 1050]$ cycles/bit, $\boldsymbol{\hat{\theta}}=[90, 280, 350, 300, 400, 1700]$ cycles/bit and $\boldsymbol{\tilde{\theta}}=[2.5296, 5.0425, 6.837, 7.8528, 8.7073, 26.3643]\times10^{-7}$ J/bit \cite{YZhang2020Nov}. Moreover, other important parameters are summarized in TABLE \ref{tab1}, where ${{\ell }_{i,s}}$ is the distance (in km) between BS $s$ and IMD $i$.
\begin{table}[]
\centering
\caption{SIMULATION PARAMETERS}
\begin{tabular}{cc}
\hline \toprule[0.5pt]
\rule{0pt}{8pt}\textbf{Parameter}  &\textbf{Value} \\
\hline \rule{0pt}{8pt}
\rule{0pt}{8pt} System bandwidth $\varpi$ & 20 MHz \\
\rule{0pt}{8pt} Noise power $\sigma^2$ & ${10}^{-11}$ mW\\
\rule{0pt}{8pt} IMD power $p_{i}^{\max}$& 23 dBm\\
\rule{0pt}{8pt} Deadline ${{\tau}_{i}^{\max}}$  & 5$\sim$10 s \\
\rule{0pt}{8pt} Data size ${{d}_{i,k}}$  & 200$\sim$500 KB \\
\rule{0pt}{8pt} Size $M$ of population & 32 \\
\rule{0pt}{8pt} Number $W$ of clusters & 5\\
\rule{0pt}{8pt} Number $K$ of tasks & 3\\
\rule{0pt}{8pt} Finance loss ${\lambda}_{k}$ & 1$\sim$5 K\$\\
\rule{0pt}{8pt} Number of SBSs at each macrocell & 30\\
\rule{0pt}{8pt} Number $L$ of cryptographic algorithms & 6\\
\rule{0pt}{8pt} Maximal security breach cost ${\psi}_{i}^{\max }$ & 5$\sim$10 K\$\\
\rule{0pt}{8pt} Security risk coefficient $\nu_{i,k}$ & 1$\sim$3\\
\rule{0pt}{8pt} Expected security level $\rho_{i,k}$ & \{5, 6\}\\
\rule{0pt}{8pt} Wired backhauling rate ${r}_{0}$  & 1 Gbps \\
\rule{0pt}{8pt} Computation capacity $f_{s}^{BS}$  & 20 GHz \\
\rule{0pt}{8pt} Computation capacity $f_{i}^{UE}$ & 1 GHz\\
\rule{0pt}{8pt} ${{c}_{i,k}}$ used for computing one bit of ${{d}_{i,k}}$ & 50$\sim$100 cycles/bit\\
\rule{0pt}{8pt} Pathloss between MBS 0 and IMD $i$ & $128.1+37.6\log_{10}\left( {{\ell }_{i,0}} \right)$\\
\rule{0pt}{8pt} Pathloss between SBS $s$ and IMD $i$ & $140.7+36.7\log_{10}\left( {{\ell }_{i,s}} \right)$\\
\rule{0pt}{8pt} Log-normal shadowing fading & Standard deviation of 8 dB\\ \bottomrule[0.5pt]
\end{tabular}
\label{tab1}
\end{table}
\par
To highlight the effectiveness of IWOA, the following algorithms are introduced for comparison.
\par
\noindent
\textbf{\textit{Computation at Mobile Terminals (CMT):}} All IMDs complete their computation tasks by themselves in allowable maximum computing capacity.
\par
\noindent
\textbf{\textit{Computation at MEC Servers (CMS):}} All computation tasks are offloaded from IMDs to BSs with the best channel gains. In addition, cryptographic algorithms with minimum security breach costs are always selected for these tasks. According to the ratio of CPU cycles used for tackling them, the computation capacities of any BS are allocated to its served tasks proportionally.
\par
\noindent
\textbf{\textit{Whale Optimization Algorithm (WOA):}} To solve the problem \eqref{eq14}, WOA in \cite{QVPham2020Apr} is introduced.
\par
In the simulation, we mainly investigate the impacts of the number of IMDs at each macrocell, and the frequency spectrum partitioning factor on the offloading performance. Due to the consideration of the historically best agent, Cauchy's inverse cumulative distribution function used for the search of prey, and the search of prey in the nearby area of the historically best agent in IWOA, IWOA may achieve lower total local energy consumption than WOA in general, and the former may also achieve higher fitness (function) value than the latter, which will be illustrated in the subsequent simulation. In addition, CMS may achieve the lowest total local energy consumption among all algorithms since it has not locally executed tasks, but CMT may achieve the highest one among all algorithms since it lets all tasks of IMDs be executed locally in allowable maximum computing capacity. As we know, in order to achieve lower local energy consumption, lower local and/or remote computation capacities may be used, resulting in higher task delay. Consequently, among all algorithms, CMS may achieve the highest total delay, CMT may have the lowest one, and IWOA may have a higher one than WOA. Under some large enough penalty factors, the latency and cost constraints of IMDs may be guaranteed strictly in WOA and IWOA. Since the support ratios of cost constraints of all algorithms are always 1, such a performance metric will not be illustrated in the following simulation, where the cost support ratio refers to the ratio of IMDs whose costs are less than or equal to the total security breach costs of them to all IMDs.
\begin{figure}[!t]
	\centering
	\centerline{\includegraphics[width=3.6in]{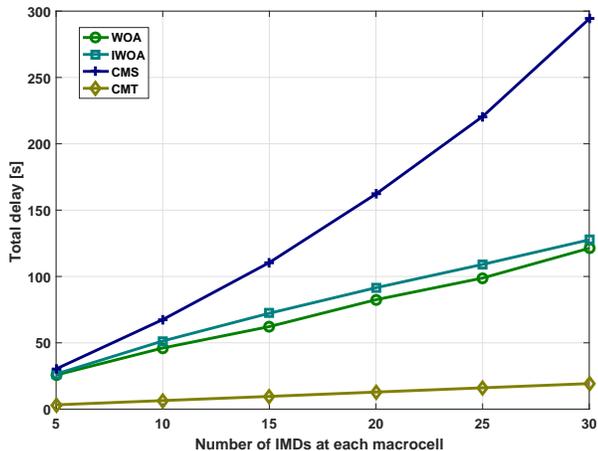}}
	\caption{Impacts of the number of IMDs at each macrocell on total task delay.}
	\label{fig3}
\end{figure}
\par
Fig.\ref{fig3} shows the impacts of the number of IMDs at each macrocell on total task delay. As illustrated in Fig.\ref{fig3}, the total task delay of all algorithms may increase with the number of IMDs at each macrocell. Such a performance trend can be easily inferred according to the definition of total task delay in \eqref{eq12}.
\begin{figure}[!t]
	\centering
	\centerline{\includegraphics[width=3.6in]{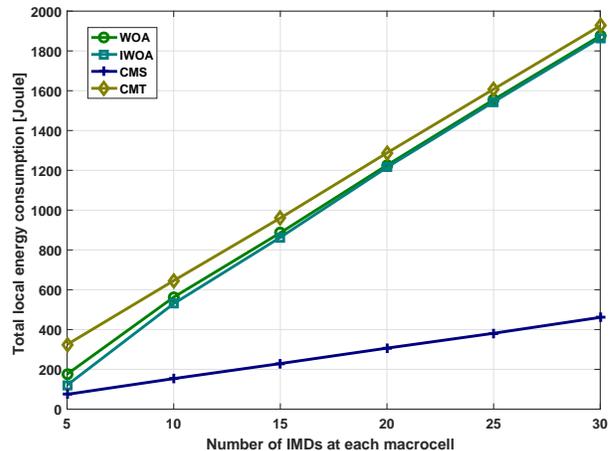}}
	\caption{Impacts of the number of IMDs at each macrocell on total local energy consumption.}
	\label{fig4}
\end{figure}
\par
Fig.\ref{fig4} shows the impacts of the number of IMDs at each macrocell on total local energy consumption. As illustrated in Fig.\ref{fig3}, the total local energy consumption of all algorithms may increase with the number of IMDs at each macrocell. Such a performance trend can be easily inferred according to the definition of total local energy consumption in \eqref{eq13}.
\begin{figure}[!t]
	\centering
	\centerline{\includegraphics[width=3.6in]{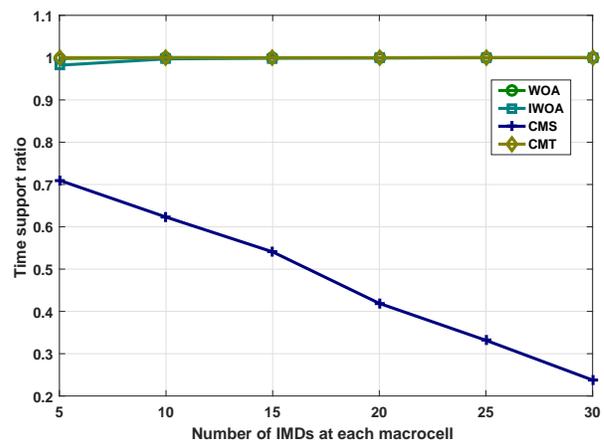}}
	\caption{Impacts of the number of IMDs at each macrocell on time support ratio.}
	\label{fig5}
\end{figure}
\par
Fig.\ref{fig5} shows the impacts of the number of IMDs at each macrocell on the time support ratio, where the mentioned ratio refers to the ratio of IMDs whose task delay is less than or equal to the deadlines of them to all IMDs. As illustrated in Fig.\ref{fig5}, WOA, IWOA and CMT almost certainly meet the latency constraints of all IMDs. Under some large enough penalty factors, the latency constraints of all IMDs in WOA and IWOA are forced to be met. Since CMT has no uplink transmission delay and encrypting delay, and it always completes the computation tasks of all IMDs in the allowable maximum computing capacity, the latency constraints of all IMDs in it can be guaranteed strictly. Unlike other algorithms, the time support ratio of CMS may decrease with the number of IMDs at each macrocell. According to the rules of CMS, we can easily know that tasks of IMDs may be always offloaded to BSs with the best channel gains. When the number of IMDs at each macrocell increases, loads of these BSs are getting heavier, resulting in the latency constraints of more and more IMDs can not be guaranteed.
\begin{figure}[!t]
	\centering
	\centerline{\includegraphics[width=3.6in]{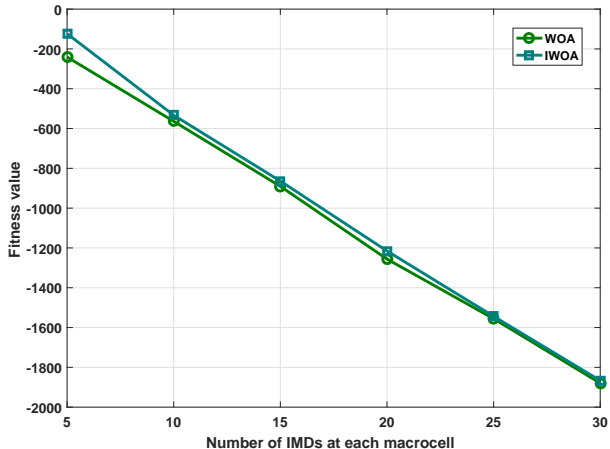}}
	\caption{Impacts of the number of IMDs at each macrocell on fitness value.}
	\label{fig6}
\end{figure}
\par
Fig.\ref{fig6} shows the impacts of the number of IMDs at each macrocell on the fitness (function) value. As illustrated in Fig.\ref{fig6}, the fitness values of WOA and IWOA may decrease with the number of IMDs at each macrocell. Such a performance trend can be easily inferred according to the definition of fitness value in \eqref{eq15}.
\begin{figure}[!t]
	\centering
	\centerline{\includegraphics[width=3.6in]{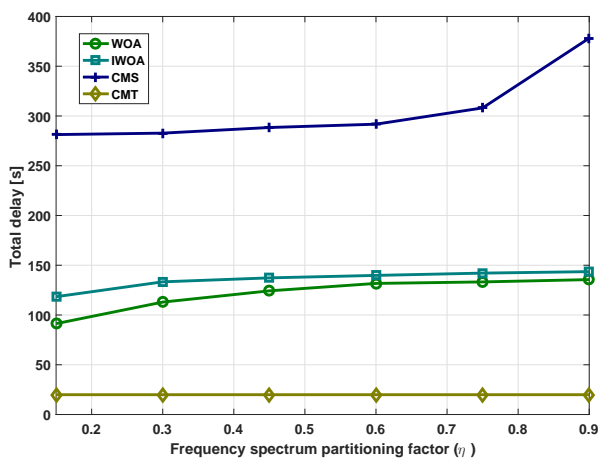}}
	\caption{Impacts of frequency spectrum partitioning factor on total task delay.}
	\label{fig7}
\end{figure}
\par
Fig.\ref{fig7} shows the impacts of frequency spectrum partitioning factor $\eta$ on total task delay. As illustrated in Fig.\ref{fig7}, besides CMT, the total task delay of other algorithms may increase with $\eta$. Since CMT doesn't utilize the uplink frequency spectrum, the total task delay of CMT should not change with $\eta$. It is easy to find that the number of NOMA channels decreases with $\eta$. Consequently, co-channel interferences may become severer and severer, resulting in increasing task delay in WOA, IWOA and CMS. Significantly, in the simulation, we find that the number of IMDs associated with SBSs in CMS is distinctly greater than the one in WOA and IWOA. It means that the total task delay of CMS may increase with $\eta$ constantly. However, the total task delay of WOA and IWOA may initially increase with $\eta$ but then doesn't change with it.
\begin{figure}[!t]
	\centering
	\centerline{\includegraphics[width=3.6in]{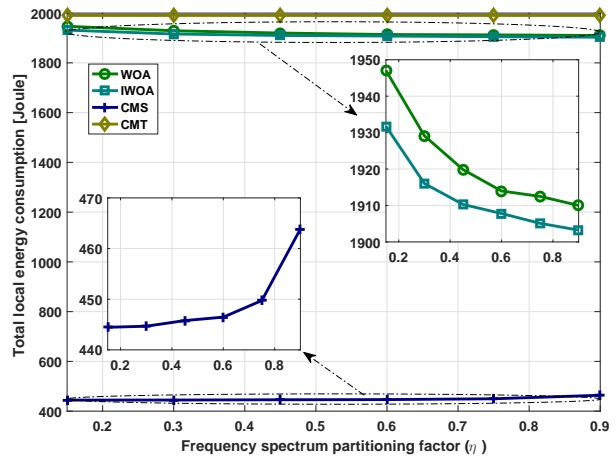}}
	\caption{Impacts of frequency spectrum partitioning factor on total local energy consumption.}
	\label{fig8}
\end{figure}
\par
Fig.\ref{fig8} shows the impacts of the frequency spectrum partitioning factor $\eta$ on total local energy consumption. As illustrated in Fig.\ref{fig8}, the total local energy consumption of CMT doesn't change with $\eta$ since it has no relation to such a factor. The total local energy consumption of WOA and IWOA may decrease with $\eta$ since the spectrum resources of MBSs selected by most IMDs increase. However, the total local energy consumption of CMS may increase with $\eta$ since the spectrum resources of SBSs selected by a lot of IMDs decrease and IMDs served by these SBSs receive severer and severer co-channel interferences. Significantly, the opposite performance trend between CMS and whale optimization algorithms may be tightly dependent on the association results of IMDs.
\begin{figure}[!t]
	\centering
	\centerline{\includegraphics[width=3.6in]{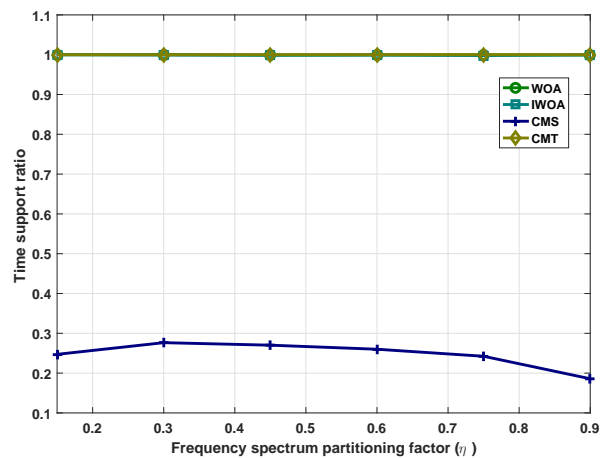}}
	\caption{Impacts of frequency spectrum partitioning factor on time support ratio.}
	\label{fig9}
\end{figure}
\par
Fig.\ref{fig9} shows the impacts of the frequency spectrum partitioning factor on the time support ratio. As illustrated in Fig.\ref{fig9}, WOA, IWOA and CMT almost certainly meet the latency constraints of all IMDs. As revealed in Fig.\ref{fig5}, the latency constraints of all IMDs in WOA, IWOA and CMT can be guaranteed strictly. Unlike other algorithms, the time support ratio of CMS may initially increase with $\eta$ but then decrease with it. In CMS, most IMDs are associated with MBSs according to the best gain association. An increased $\eta$ results in increased uplink data rates of IMDs associated with MBSs, resulting in an increased time support ratio. However, an increased $\eta$ also results in decreased uplink data rates of IMDs associated with SBSs because of fewer spectrum resources and severer co-channel interferences. It may result in a decreased time support ratio.
\begin{figure}[!t]
	\centering
	\centerline{\includegraphics[width=3.6in]{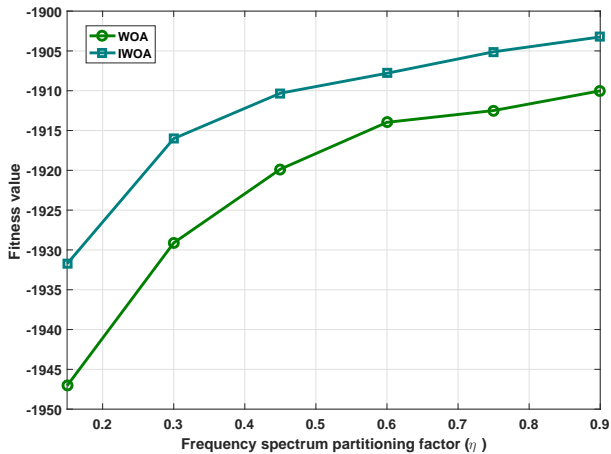}}
	\caption{Impacts of frequency spectrum partitioning factor on fitness value.}
	\label{fig10}
\end{figure}
\par
Fig.\ref{fig10} shows the impacts of the frequency spectrum partitioning factor on the fitness (function) value. As illustrated in Fig.\ref{fig10}, the fitness values of WOA and IWOA may increase with $\eta$. Seen from Fig.\ref{fig8}, the total local energy consumption of WOA and IWOA decreases with $\eta$. According to the definition of fitness value in \eqref{eq15}, we can easily know that decreased total local energy consumption may result in increased fitness value.

\begin{figure}[!t]
	\centering
	\centerline{\includegraphics[width=3.6in]{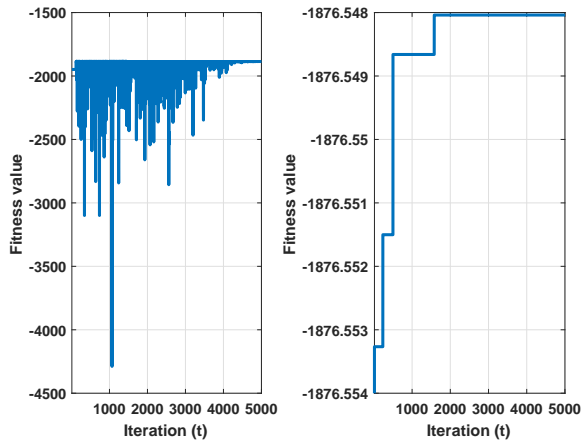}}
	\caption{Convergence comparison of WOA and IWOA.}
	\label{fig11}
\end{figure}
\par
Fig.\ref{fig11} shows the convergence of WOA and IWOA. As illustrated in Fig.\ref{fig11}, IWOA has a higher convergence rate than WOA. In addition, the former can achieve higher fitness value than the latter. Evidently, by considering the historically best agent, Cauchy's inverse cumulative distribution function used for the search of prey, and the search of prey in the nearby area of the historically best agent, IWOA may achieve better performance than WOA.
\section{CONCLUSION}\label{sec 7}
As for ultra-dense multi-task IoT networks, both OMA and NOMA are first used to mitigate network interferences and improve spectrum utilization. Then, under the proportional allocation of computational resources and the constraints of latency and security cost, we jointly optimize device association, channel selection, security service assignment, power control and multi-step computation offloading to minimize the total energy consumption of all IMDs. Considering that the finally formulated problem is in a nonlinear mixed-integer form and hard to tackle, we design IWOA to solve it. After that, the convergence, computational complexity and parallel implementation are analyzed in detail. Simulation results show that IWOA may achieve lower energy consumption than other existing algorithms under the constraints of latency and security cost. Future work can include further improvement of IWOA, and the application of data compression and other intelligent algorithms.

\end{document}